\documentclass{jetpl}

\usepackage{graphicx}
\usepackage{cite}

\twocolumn
\lat

\DeclareMathOperator{\sgn}{sgn}

 \title{Triplet proximity effect in FSF trilayers}
 \rtitle{Triplet proximity effect in FSF trilayers}
 \sodtitle{Triplet proximity effect in FSF trilayers}

 \author{Ya.\,V.~Fominov$^{\,+\,*}$\/\thanks{e-mail: fominov@landau.ac.ru, a.golubov@tn.utwente.nl,
                                                     mkupr@pn.sinp.msu.ru},
         A.\,A.~Golubov$^{\,*\,1)}$,
         M.\,Yu.~Kupriyanov$^{\,\square\,1)}$}
 \rauthor{Ya.\,V.~Fominov, A.\,A.~Golubov, and M.\,Yu.~Kupriyanov}
 \sodauthor{Fominov, Golubov, and Kupriyanov}

\address{
$^+$ L.\,D.~Landau Institute for Theoretical Physics RAS, 119334 Moscow, Russia\\
$^*$ Department of Applied Physics, University of Twente, 7500 AE Enschede, The Netherlands\\
$^\square$ Nuclear Physics Institute, Moscow State University, 119992 Moscow, Russia}

\dates{3 July 2003}{*}

\abstract{We study the critical temperature $T_c$ of FSF trilayers (F is a ferromagnet, S is a singlet superconductor),
where the triplet superconducting component is generated at noncollinear magnetizations of the F layers. An exact
numerical method is employed to calculate $T_c$ as a function of the trilayer parameters, in particular, mutual
orientation of magnetizations. Analytically, we consider limiting cases. Our results determine conditions which are
necessary for existence of recently investigated odd triplet superconductivity in SF multilayers.}


\begin{document}
\maketitle

A striking feature of the proximity effect between singlet superconductors and nonhomogeneous ferromagnets is the
possibility of generating the triplet superconducting component \cite{Bergeret,Kadigrobov}. Recently, it was shown that
the triplet component also arises in the case of several homogeneous but differently oriented ferromagnets \cite{VBE}.
Physically, the generating of the triplet component in SF systems \cite{Bergeret,Kadigrobov,VBE} is similar to the case
of magnetic superconductors \cite{JLTP}.

In Ref. \cite{VBE}, the Josephson effect was studied having in mind that the superconductivity in the system is not
suppressed by the ferromagnets. However, this issue requires separate study.

Although the SF proximity effect is rather well studied, the influence of the \textit{mutual orientation} of F layers
magnetizations (exchange fields) on $T_c$ of layered SF structures has been mostly considered basing on the cases of
parallel (P) and antiparallel (AP) alignment \cite{Tagirov_PRL,Buzdin_EL,Khusainov,Buzdin_cm,Deutscher,ANL}. At the same
time, those are the only cases when the triplet component is absent.

A FSF trilayer with homogeneous but noncollinear magnetizations of the F layers is the simplest example of a layered
structure in which the triplet component is generated. The triplet component (correlations between quasiparticles with
parallel spins) arises as a result of interplay between the Andreev reflections at the two SF interfaces. This mechanism
is similar to the one described in Ref. \cite{Kadigrobov}, with the difference that instead of local magnetic
inhomogeneity we deal with magnetic inhomogeneity of the structure as a whole.

The critical temperature of the noncollinear FSF system was studied in Ref. \cite{Buzdin}. However, in that work the
triplet component was not taken into account. Thus calculation of $T_c$ in the noncollinear FSF trilayer is still an
open question.

\begin{figure}[tb]
 \centerline{\includegraphics[width=78mm]{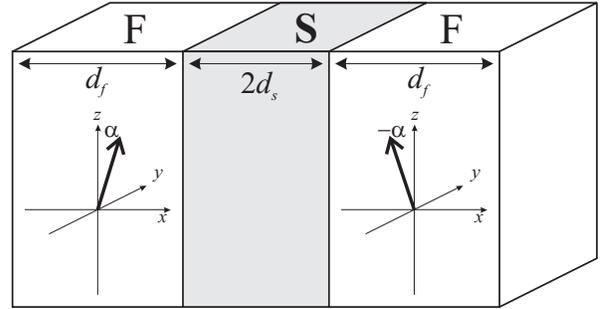}}
\caption{Fig.\protect\ref{fig:FSF}. FSF trilayer. The system is the same as in Ref. \cite{Buzdin}. The thickness of the
S layer is $2d_s$, of each F layer --- $d_f$. The center of the S layer corresponds to $x=0$. The thick arrows in the F
layers denote the exchange fields $\mathbf{h}$ lying in the $(y,z)$ plane. The angle between the in-plane exchange
fields is $2\alpha$.}
 \label{fig:FSF}
\end{figure}

In this letter we study the critical temperature of a FSF trilayer at arbitrary angle between the in-plane
magnetizations (see Fig.\ref{fig:FSF}), which makes it necessary to take the triplet component into account. We reduce
the problem to the form, which allows to apply general numerical methods developed in Refs. \cite{JETPL,PRB}. This form
also leads to some general conclusions about $T_c$ and allows analytical progress in limiting cases.

\textbf{1. General description.} We consider the dirty limit, which is described by the Usadel equations. Near $T_c$,
the Usadel equations are linearized and contain only the anomalous Green function $\widehat F$ \cite{Bergeret}:
\begin{gather}
\frac D2 \frac{d^2 \widehat F}{dx^2} - |\omega_n| \widehat F +\Delta \widehat \sigma_3 -\frac i2
\sgn\omega_n \left( \widehat F \widehat H^* +\widehat H \widehat F \right) =0, \notag \\
\widehat F = \begin{pmatrix} f_{\uparrow\downarrow} & f_{\uparrow\uparrow} \\ f_{\downarrow\downarrow} &
f_{\downarrow\uparrow} \end{pmatrix} . \label{Usadel}
\end{gather}
Here $D$ is the diffusion constant ($D_s$ and $D_f$ for the S and F layers), $\omega_n=\pi T(2n+1)$ are the Matsubara
frequencies, and $\widehat \sigma_3$ is the third Pauli matrix. The function $\widehat F$ is a matrix in the spin space.
The $f_{\uparrow\uparrow}$ and $f_{\downarrow\downarrow}$ components describe the triplet superconducting correlations.
In the P and AP cases it is sufficient to consider only the scalar equation for the singlet component
$f_{\uparrow\downarrow}$.

Equation (\ref{Usadel}) is written in the general case when both pair potential and exchange field are present. In our
system, in the F layers the pair potential is absent, $\Delta=0$, while
\begin{equation}
\widehat H = h \left( \widehat \sigma_2 \sin\alpha + \widehat \sigma_3 \cos\alpha \right)
\end{equation}
at the exchange field $\mathbf{h} = h(0,\sin\alpha,\cos\alpha)$. $h$ is the exchange energy, and $\alpha$ describes the
direction of the in-plane magnetization.

In the S layer, the exchange energy is zero, while the pair potential obeys the self-consistency equation
\begin{equation}
\Delta\ln\frac{T_{cs}}{T} = \pi T \sum_{\omega_n} \left( \frac\Delta{|\omega_n|} -f_{\uparrow\downarrow} \right),
\end{equation}
where $T_{cs}$ is the critical temperature of the S material. In the case of a single S layer, $\Delta$ can be chosen
real.

The boundary conditions at the outer surfaces of the trilayer are
\begin{equation} \label{vacuum}
d \widehat F_f / dx = 0,
\end{equation}
while at the SF interfaces
\begin{align}
& \xi_s( d \widehat F_s / dx ) =\gamma \xi_f ( d \widehat F_f / dx ), &&
\gamma = \rho_s \xi_s / \rho_f \xi_f, \label{bound_1} \\
& \pm \xi_f \gamma_b ( d \widehat F_f / dx) = \widehat F_s -\widehat F_f, && \gamma_b = R_b {\cal A} / \rho_f \xi_f.
\label{bound_2}
\end{align}
Here $\xi_{s(f)}=\sqrt{D_{s(f)}/2\pi T_{cs}}$ and $\rho_{s(f)}$ are the coherence lengths and the normal state
resistivities of the S and F metals, $R_b$ is the total resistance of the SF boundary, and $\cal A$ is its area. The
$\pm$ sign in the l.h.s. of Eq. (\ref{bound_2}) refers to the left and right SF interface, respectively. The above
boundary conditions were derived for SN interfaces \cite{KL} (N is a normal metal); their use in the SF case is
justified by the small parameter $h/E_F \ll 1$ ($E_F$ is the Fermi energy).

%Our strategy is to reduce the problem to the S layer only, with effective boundary conditions.

We expand the Green function $\widehat F$ in the basis of the Pauli matrices $\widehat \sigma_i$, $i=1,2,3$, and the
unity matrix $\widehat\sigma_0$. It can be shown that the solution has the form
\begin{equation}
\widehat F = f_0 \widehat\sigma_0 + f_1 \widehat\sigma_1 + f_3 \widehat\sigma_3.
\end{equation}
The $f_0$ component is imaginary, while $f_1$ and $f_3$ are real. The relations $f_0(-\omega_n)=-f_0(\omega_n)$,
$f_1(-\omega_n)=-f_1(\omega_n)$, $f_3(-\omega_n)=f_3(\omega_n)$ make it sufficient to consider only positive Matsubara
frequencies.

The $f_1$ component describes a special type of triplet condensate \cite{Bergeret,VBE}, odd in frequency
[$f_1(-\omega_n)=-f_1(\omega_n)$] and even in momentum, which is similar to the one proposed by Berezinskii
\cite{Berezinskii}. It is independence on the momentum direction that allows the triplet condensate to survive in the
diffusive limit.
%, in contrast to the standard (odd in momentum) case \cite{Larkin}.

Equation (\ref{Usadel}) yields three coupled scalar equations (we consider $\omega_n>0$):
\begin{gather}
\frac D2 \frac{d^2 f_0}{dx^2} - \omega_n f_0 -i h f_3 \cos\alpha =0, \notag \\
\frac D2 \frac{d^2 f_1}{dx^2} - \omega_n f_1 + h f_3 \sin\alpha =0, \label{sys_gen} \\
\frac D2 \frac{d^2 f_3}{dx^2} - \omega_n f_3 -i h f_0 \cos\alpha -h f_1 \sin\alpha +\Delta =0. \notag
\end{gather}
Analyzing symmetries implied by Eqs. (\ref{sys_gen}) and geometry of the system, we conclude that $f_0(x)=f_0(-x)$,
$f_1(x)=-f_1(-x)$, $f_3(x)=f_3(-x)$. Thus we can consider only one half of the system, say $x<0$, while the boundary
conditions at $x=0$ are
\begin{equation} \label{bound_components}
d f_0 / dx = 0,\qquad f_1 =0,\qquad d f_3 / dx =0.
\end{equation}

Below we shall use the following wave vectors:
\begin{gather}
k_f = \sqrt{2\omega_n / D_f},\qquad k_h = \sqrt{h / D_f}, \notag \\
\widetilde k_h = \sqrt{k_f^2 + 2i k_h^2},\qquad k_s = \sqrt{2\omega_n / D_s}. \label{k}
\end{gather}
The solution in the left F layer, satisfying the boundary condition (\ref{vacuum}), has the form
\begin{gather}
\widehat F_f = C_1 \left( i \widehat\sigma_0 \sin\alpha +\widehat\sigma_1 \cos\alpha
\right) \cosh\left[ k_f \left( x+d_s +d_f \right) \right] + \notag \\
+ C_2 \left( \widehat\sigma_0 \cos\alpha + i\widehat\sigma_1 \sin\alpha +\widehat\sigma_3
\right) \cosh\left[ \widetilde k_h \left( x+d_s +d_f \right) \right] + \notag \\
+ C_3 \left( \widehat\sigma_0\cos\alpha  + i\widehat\sigma_1 \sin\alpha -\widehat\sigma_3 \right) \cosh\left[ \widetilde
k_h^* \left( x+d_s +d_f \right) \right] .
\end{gather}
The matrix boundary condition (\ref{bound_2}) yields three scalar equations, which allow to express the coefficients
$C_1$, $C_2$, $C_3$ in terms of the components $f_0$, $f_1$, $f_3$ of the Green function on the S side of the FS
interface:
\begin{align}
C_1 &= \left( -i f_0 \sin\alpha+ f_1 \cos\alpha \right) / \left( 1+\gamma_b A_f \right) , \notag \\
C_2 &= \left( f_0 \cos\alpha -i f_1 \sin\alpha +f_3 \right) / 2 \left( 1+\gamma_b A_h \right), \\
C_3 &= \left( f_0 \cos\alpha -i f_1 \sin\alpha -f_3 \right) / 2 \left( 1+\gamma_b A_h^* \right), \notag
\end{align}
where we have introduced the following notations:
\begin{gather}
A_f = k_f \xi_f \tanh (k_f d_f), \quad
A_h = \widetilde k_h \xi_f \tanh (\widetilde k_h d_f), \notag \\
V_f = \gamma A_f / (1+\gamma_b A_f), \quad V_h = \gamma A_h / (1+\gamma_b A_h). \label{AO}
\end{gather}
Then the boundary condition (\ref{bound_1}) yields
%\begin{gather}
%\xi_s ( d f_0 / dx) = f_0 \left( V_f \sin^2 \alpha + \Real V_h   \cos^2 \alpha \right) - \notag \\
%-i f_1 \left( V -\Real V_h \right) \sin\alpha \cos\alpha + i f_3 \Imag V_h \cos\alpha, \label{f_0} \\
%\xi_s ( d f_1 / dx) = i f_0 \left( V_f - \Real V_h \right) \sin\alpha \cos\alpha + \notag \\
%+f_1 \left( V_f \cos^2 \alpha + \Real V_h \sin^2 \alpha \right) + f_3 \Imag V_h \sin\alpha, \\
%\xi_s ( d f_3 / dx) = if_0\Imag V_h \cos\alpha - \notag \\
%-f_1 \Imag V_h \sin\alpha + f_3 \Real V_h. \label{f_3}
%\end{gather}
three scalar equations which entangle $f_0$, $f_1$, and $f_3$. Thus the Green function of the F layer is eliminated, and
we obtain equations for the S layer only. Moreover, we can proceed further, because in the S layer the unknown
function $\Delta(x)$ %(which must be determined self-consistently)
only enters the equation for the $f_3$ component [see
Eqs. (\ref{sys_gen})]. At the same time, taking boundary conditions (\ref{bound_components}) into account, we can write
$f_0 = B_0 \cosh (k_s x)$, $f_1 = B_1 \sinh (k_s x)$. Excluding $B_0$ and $B_1$, we arrive at the effective boundary
condition for $f_3$:
\begin{equation}
\xi_s ( d f_3 / dx) = W f_3,
\end{equation}
where
\begin{equation} \label{W1}
W = \Real V_h + \frac{\left( \Imag V_h \right)^2} {k_s \xi_s A(\alpha) +\Real V_h},
\end{equation}
and the angular dependence is determined by
\begin{equation} \label{A}
A = \frac{k_s \xi_s \tanh (k_s d_s) +V_f \left[ \sin^2 \alpha +\tanh^2 (k_s d_s) \cos^2 \alpha \right]} {k_s \xi_s
\left[ \cos^2 \alpha +\tanh^2 (k_s d_s) \sin^2 \alpha \right] +V_f \tanh (k_s d_s)}.
\end{equation}

Effectively, we obtain the following problem:
\begin{gather}
\Delta \ln \frac{T_{cs}}T = 2\pi T \sum_{\omega_n>0} \left( \frac\Delta{\omega_n} - f_3 \right), \label{1} \\
\frac{D_s}2 \frac{d^2 f_3}{dx^2} - \omega_n f_3 +\Delta =0, \label{2} \\
\xi_s \frac{d f_3(-d_s)}{dx} = W(\omega_n) f_3(-d_s),\qquad \frac{d f_3(0)}{dx} =0 \label{3}
\end{gather}
--- this is exactly the problem that was solved in Refs. \cite{JETPL,PRB}. Inserting the new function
$W$, we can use the methods developed in those works. At $\alpha=0$, Eq. (\ref{W1}) reproduces $W$ from Refs.
\cite{JETPL,PRB}.

All information about the F layers is contained in a single function $W$, all information about the misorientation angle
--- in its part $A(\alpha)$. Knowledge of $W$ is already sufficient to draw several general conclusions about the
behavior of $T_c$. First, if the S layer is thick, i.e. $d_s \gg \xi_s$, then $\tanh (k_s d_s) \approx 1$ at
characteristic frequencies, and $T_c$ does not depend on $\alpha$. Qualitatively, this happens because the effect of
mutual orientation of the F layers is due to ``interaction'' between the two SF interfaces, which is efficient only in
the case of thin S layer. Second, $T_c$ does not depend on $d_f$ if $d_f \gg \xi_f$. Qualitatively, this is due to the
fact that the superconducting correlations penetrate from the S to F layer only on the scale $\xi_f$.
%largest of the two characteristic lengths, and if $d_f$ is even larger, it does not affect the proximity effect.

The triplet component is ``nonmonotonic'' as a function of $\alpha$: it vanishes at $\alpha=0$ and $\alpha=\pi/2$ (P and
AP case, respectively), and arises only between the two boundary values. However, the $T_c(\alpha)$ dependence is always
monotonic. It can be directly proven from the monotonic behavior of $A(\alpha)$, and, hence, $W$. This rigorously
derived conclusion disproves the result obtained by the approximate single-mode method in Ref. \cite{Khusainov}, where
it was claimed that $T_c$ in the AP configuration can be smaller than in the P case.

\begin{figure}[tb]
 \centerline{\includegraphics[width=81mm]{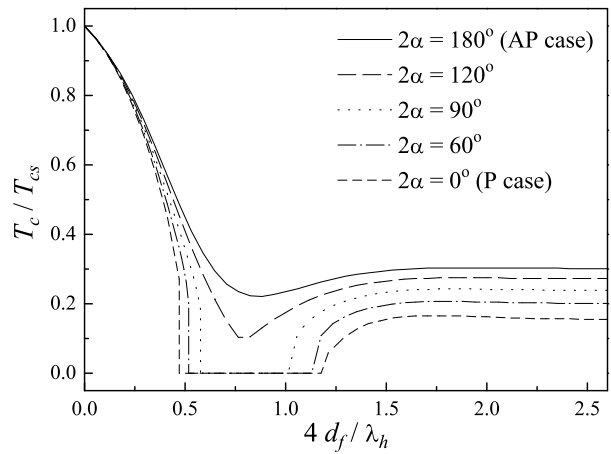}}
\caption{Fig.\protect\ref{fig:Tc_dF}. Critical temperature $T_c$ vs. thickness of the F layers $d_f$, which is
normalized on the wavelength of the singlet component oscillations $\lambda_h = 2\pi/k_h$. Parameters $d_s /\xi_s =1.2$,
$h/\pi T_{cs} =6.8$, $\gamma=0.15$, $\gamma_b=0.02$ correspond to Ref. \cite{PRB}. The curves are calculated at
different angles $2\alpha$ between the in-plane exchange fields in the F layers.}
 \label{fig:Tc_dF}
\end{figure}

\begin{figure}[tb]
 \centerline{\includegraphics[width=84mm]{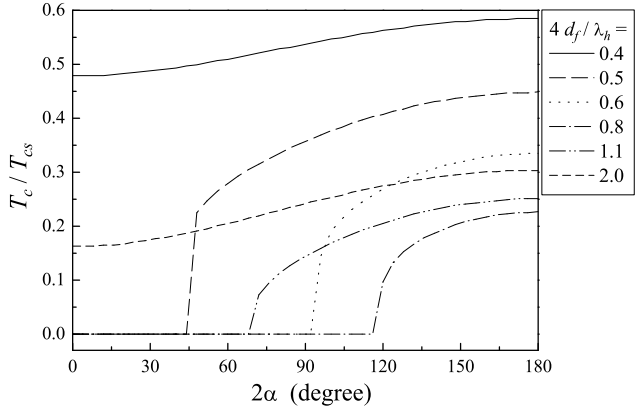}}
\caption{Fig.\protect\ref{fig:Tc_alpha}. $T_c$ vs. misorientation angle $2\alpha$. The curves correspond to different
thicknesses of the F layers $d_f$. The parameters are the same as in Fig.\ref{fig:Tc_dF}.}
 \label{fig:Tc_alpha}
\end{figure}

Numerical results obtained by the methods developed in Refs. \cite{JETPL,PRB}, are shown in
Figs.\ref{fig:Tc_dF},\ref{fig:Tc_alpha}. A question arises: why is there pronounced angular dependence in the case $d_s>
\xi_s$, when the S layer is not thin? The answer is that the condition $d_s \ll \xi_s = \sqrt{D_s /2\pi T_{cs}}$ is a
\textit{sufficient} condition of thin S layer, whereas the \textit{necessary} condition is weaker: $d_s \ll
\xi=\sqrt{D_s /2\pi T_c}$, since the characteristic energy for a particular system is $\pi T_c$ with its own value of
$T_c$. The two conditions become essentially different if $T_c$ is notably suppressed, and in this case $T_c$ can
exhibit pronounced angular dependence at $d_s \ll \xi$, while it is possible to have $d_s > \xi_s$.

Experimentally, the conditions for observing the angular dependence of $T_c$ are more easily met when $T_c$ is
essentially (but not completely) suppressed. Accordingly, the effect of $\alpha$ on $T_c(d_f)$ dependence is most
pronounced near the reentrant behavior. Experimental detection of such behavior was reported in Ref. \cite{reentrant}.

\textbf{2. Thin S layer.} If $d_s \ll \xi_s$, then $\Delta$ is constant. The Usadel equation (\ref{2}) can be solved,
and the equation determining $T_c$ takes the form
\begin{equation} \label{ln_thin}
\ln \frac{T_{cs}}{T_c} = 2\pi T_c \sum_{\omega_n>0} \left( \frac 1{\omega_n} - \frac 1{\omega_n +W \pi T_{cs} \xi_s/d_s}
\right),
\end{equation}
where $W$ is given by Eq. (\ref{W1}) with simplified function $A(\alpha)$:
\begin{equation} \label{A_simple}
A = \frac{k_s^2 \xi_s d_s +V_f \left[ \sin^2 \alpha +(k_s d_s)^2 \cos^2 \alpha \right]} {k_s \xi_s \left[ \cos^2 \alpha
+(k_s d_s)^2 \sin^2 \alpha \right] +V_f k_s d_s}.
\end{equation}

For the P and AP alignments, under additional assumption of strong ferromagnetism ($h\gg \pi T_{cs}$), we obtain:
\begin{gather}
\ln\frac{T_{cs}}{T_c^P} = \Real \psi \left( \frac 12 +\frac{V_h}2 \frac{\xi_s}{d_s}
\frac{T_{cs}}{T_c^P} \right) - \psi \left( \frac 12 \right), \label{eq_parall} \\
\ln\frac{T_{cs}}{T_c^{AP}} = \psi \left( \frac 12 +\frac W2 \frac{\xi_s}{d_s} \frac{T_{cs}}{T_c^{AP}} \right) - \psi
\left( \frac 12 \right), \label{eq_anti}
\end{gather}
where $\psi$ is the digamma function, $V_h$ is determined by Eqs. (\ref{AO}) with $\widetilde k_h =(1+i) k_h$, and in
the region of parameters, where $T_c \ne 0$ [the corresponding conditions can be extracted from the results for the
critical thickness --- see Eqs. (\ref{dscr}), (\ref{dscr_cond}) below], we may write
\begin{equation} \label{Wb}
W = \Real V_h + \left( \Imag V_h \right)^2  d_s / \xi_s.
\end{equation}
Due to symmetry, the result for the P case (\ref{eq_parall}) reproduces that for the SF bilayer with S layer of
thickness $d_s$ \cite{PRB}. In the AP case, if the second terms in the r.h.s. of Eq. (\ref{Wb}) can be neglected (e.g.,
at $k_h d_f \gg 1$ in the region of parameters where $T_c \ne 0$), then $W = \Real V_h$ and we reproduce the result of
Ref. \cite{Buzdin_cm}. However, the second term becomes essential in the Cooper limit, defined by conditions $d_s \ll
\sqrt{D_s / 2 \omega_D}$, $d_f \ll \min( \sqrt{D_f /2 \omega_D}, k_h^{-1} )$, $\gamma_b =0$, with $\omega_D$ the Debye
energy of the S material. In this case $\Real V_h=0$ and Eqs. (\ref{eq_anti}), (\ref{Wb}) reproduce the result of
Tagirov \cite{Tagirov_PRL}.

The critical thickness $d_{sc}$ of the S layer, below which the superconductivity vanishes, immediately follows from
Eqs. (\ref{eq_parall}), (\ref{eq_anti}) for the P and AP cases:
\begin{equation} \label{dscr}
d_{sc}^P / \xi_s = 2e^C \left| V_h \right|,\qquad d_{sc}^{AP} / \xi_s = 2e^C W
\end{equation}
at
\begin{equation} \label{dscr_cond}
d_{sc} /\xi_s \ll 1.
\end{equation}
Here $C \approx 0.577$ is Euler's constant. Condition (\ref{dscr_cond}) is necessary for applicability of Eqs.
(\ref{dscr}). If this condition is not satisfied, then Eqs. (\ref{dscr}) only tell us that at $d_s /\xi_s \ll 1$ the
superconductivity is certainly absent, i.e., $T_c=0$. According to the monotonic growth of $T_c(\alpha)$, the function
$d_{sc} (\alpha)$ decreases monotonically, hence $d_{sc}^P > d_{sc}^{AP}$. At $\gamma_b=0$, $k_h d_f \gg 1$, Eqs.
(\ref{dscr}) reproduce the results of Ref. \cite{Buzdin} for the P and AP cases.

The $T_c(\alpha)$ dependence can be most easily studied in the Cooper limit. In this case a simple analysis (see, e.g.,
Appendix A1 in Ref. \cite{PRB}) can be done already on the level of the Usadel equations, and the system is described as
a uniform layer with the effective exchange energy\footnote{Since $\omega_n$ was neglected in comparison with $h$ in the
Usadel equation, the result of the Cooper limit is valid only at $\tau_s \gg \tau_f$.}
\begin{equation} \label{h_eff}
h_\mathrm{eff} = (\tau_f / \tau_s) h \cos\alpha ,
\end{equation}
where $\tau_{s(f)} = 2 d_{s(f)} R_b \mathcal{A} / \rho_{s(f)} D_{s(f)}$. The accuracy of this result is limited to the
first order over $h$, which becomes insufficient in the vicinity of $\alpha=\pi/2$. At $\alpha =\pi/2$, the first-order
effect of $h$ vanishes, while a more accurate analysis (Ref. \cite{Tagirov_PRL} and Eqs. (\ref{eq_anti}), (\ref{Wb}))
reveals the second-order effect of $h$ on $T_c$.

Let us now consider the same limit as in  Ref. \cite{Buzdin}:
\begin{gather}
d_s \ll \xi_s,\quad k_h d_f \gg 1,\quad h \gg \pi T_{cs},\quad \gamma_b =0, \\
\gamma k_h \xi_f d_s /\xi_s \ll 1. \label{conds}
\end{gather}
The condition to have superconductivity at least at some orientations has the form $d_{sc}^{AP} < d_s \ll \xi_s$, and in
the case under discussion, Eqs. (\ref{dscr}), (\ref{dscr_cond}) yield:
\begin{equation} \label{cond_S}
2 e^C \gamma k_h \xi_f < d_s / \xi_s \ll 1,
\end{equation}
hence condition (\ref{conds}) becomes redundant.

Starting from Eqs. (\ref{ln_thin}), (\ref{W1}), (\ref{A_simple}), we finally obtain the following equation for $T_c$:
\begin{equation} \label{ln_final}
\ln \frac{T_{cs}}{T_c} = Q\, \psi \left( \frac 12 +\frac{\Omega_1}{2\pi T_c} \right) +R\, \psi \left( \frac 12
+\frac{\Omega_2}{2\pi T_c} \right) -\psi \left( \frac 12 \right),
\end{equation}
where
\begin{gather}
Q = \frac 12 +\frac{\sin^2 \alpha}{2\sqrt{\sin^4 \alpha -4 \cos^2 \alpha}},\qquad R = 1-Q, \notag \\
\Omega_{1,2} = \frac{d_0}{d_s} \pi T_{cs} \left( 1+\cos^2 \alpha \pm \sqrt{\sin^4 \alpha -4 \cos^2 \alpha} \right),
\notag \\
d_0 = \gamma k_h \xi_f \xi_s /2. \label{legend1}
\end{gather}
In the P and AP cases, where the triplet component is absent, Eqs. (\ref{ln_final}), (\ref{legend1}) reproduce the
results of Refs. \cite{Buzdin_EL,Buzdin}. At the same time, at a noncollinear alignment the results are clearly
different.

\begin{figure}[tb]
 \centerline{\includegraphics[width=71mm]{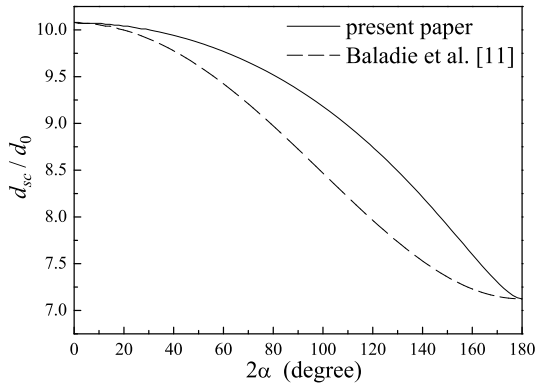}}
\caption{Fig.\protect\ref{fig:d_sc}. Critical thickness of the S layer $d_{sc}$ vs. misorientation angle $2\alpha$.
Dashed line is the result of Ref. \cite{Buzdin}, obtained without account of the triplet component.}
 \label{fig:d_sc}
\end{figure}

The critical thickness is found from Eqs. (\ref{ln_final}), (\ref{legend1}):
\begin{gather}
d_{sc} (\alpha) / d_0 = 4\sqrt{2} e^C \cos\alpha \times \label{dscr_alpha} \\
\times \left( \frac{1+\cos^2 \alpha + \sqrt{\sin^4 \alpha -4 \cos^2 \alpha}} {1+\cos^2 \alpha - \sqrt{\sin^4 \alpha -4
\cos^2 \alpha}} \right)^ {\textstyle\frac{\sin^2 \alpha} {2 \sqrt{\sin^4 \alpha -4 \cos^2 \alpha}}}. \notag
\end{gather}
Although the square root in this expression can become imaginary, the whole expression remains real ($z^i$ is real if
$|z|=1$). Figure \ref{fig:d_sc} illustrates the result (\ref{dscr_alpha}).

Now we turn to analyze the conditions of applicability for the results reported in Ref. \cite{VBE}. A noncollinear FSF
trilayer is a unit cell of the multilayered structure studied in that work. The main result of Ref. \cite{VBE}, the
Josephson current due to the long-range triplet component, requires that the S layer is thin $d_s \ll \xi_s$, while the
F layers are thick for the singlet component and moderate for the triplet one: $k_h^{-1} \ll \xi_f < d_f$ \cite{VBE}. In
this case the condition that superconductivity is not completely suppressed at least in the vicinity of the AP alignment
[Eqs.(\ref{dscr}),(\ref{dscr_cond})] takes the form
\begin{equation} \label{cond_S1}
4 e^C \gamma k_h \xi_f \frac{1+2 \gamma_b k_h \xi_f}{(1+2 \gamma_b k_h \xi_f)^2 +1} < \frac{d_s}{\xi_s} \ll 1.
\end{equation}
At $\gamma_b=0$ (as it was assumed in Ref. \cite{VBE}), this yields $2 e^C \gamma k_h \xi_f < d_s /\xi_s \ll 1$, which
is a rather strong condition for $\gamma$, since $k_h \xi_f \gg 1$. Finite interface transparency relaxes this
condition: already at $\gamma_b \sim 1$, Eq. (\ref{cond_S1}) yields $2 e^C \gamma / \gamma_b < d_s /\xi_s \ll 1$.

The condition that superconductivity exists at all orientations has the form similar to Eq. (\ref{cond_S1}) but with the
corresponding expression for $d_{sc}^P$ instead of $d_{sc}^{AP}$ in the l.h.s. This only leads to a minor difference,
since the two critical thicknesses are of the same order: $d_{sc}^P = \sqrt{2} d_{sc}^{AP}$ at $\gamma_b =0$, while
$d_{sc}^P = d_{sc}^{AP}$ at $\gamma_b > 1$.

In conclusion, we have studied $T_c$ of a FSF trilayer as a function of its parameters, in particular, the angle between
magnetizations of the F layers. The angular dependence becomes pronounced when the S layer is thin, and can lead to
switching between superconducting and non-superconducting states as the angle is varied. Our results directly apply to
multilayered SF structures, where a FSF trilayer is a unit cell. We have formulated the conditions which are necessary
for existence of recently investigated odd triplet superconductivity in SF multilayers \cite{VBE}.

We thank M.\,V.~Feigel'man and N.\,M.~Chtchel\-katchev for discussions. The research was supported by the ESF PiShift
program. Ya.V.F. was also supported by the RFBR 01-02-17759, the Swiss NF, the Russian Ministry of Industry, Science and
Technology (RMIST), and the program ``Quantum Macrophysics'' of the RAS. A.A.G. was also supported by the INTAS-01-0809.
M.Yu.K. was also supported by the RMIST.


\begin{thebibliography}{99}

\bibitem{Bergeret}
F.\,S.~Bergeret, K.\,B.~Efetov, and A.\,I.~Larkin, Phys. Rev. \textbf{B62}, 11872 (2000); F.\,S.~Bergeret,
A.\,F.~Volkov, and K.\,B.~Efetov, Phys. Rev. Lett. \textbf{86}, 4096 (2001).

\bibitem{Kadigrobov}
A.~Kadigrobov, R.\,I.~Shekhter, and M.~Jonson, Europhys. Lett. \textbf{54}, 394 (2001); Fiz. Nizk. Temp.
\textbf{27}, 1030 (2001) [Low Temp. Phys. \textbf{27}, 760 (2001)].

\bibitem{VBE}
A.\,F.~Volkov, F.\,S.~Bergeret, and K.\,B.~Efetov, Phys. Rev. Lett. \textbf{90}, 117006 (2003).

\bibitem{JLTP}
L.\,N.~Bulaevskii, A.\,I.~Rusinov, and M.~Kuli\'c, J. Low Temp. Phys. \textbf{39}, 255 (1980).

\bibitem{Tagirov_PRL}
L.\,R.~Tagirov, Phys. Rev. Lett. \textbf{83}, 2058 (1999).

\bibitem{Buzdin_EL}
A.\,I.~Buzdin, A.\,V.~Vedyayev, and N.\,V.~Ryzhanova, Europhys. Lett. \textbf{48}, 686 (1999).

\bibitem{Khusainov}
M.\,G.~Khusainov, Yu.\,A.~Izyumov, and Yu.\,N.~Proshin, Pis'ma Zh. Eksp. Teor. Fiz. \textbf{73}, 386
(2001) [JETP Lett. \textbf{73}, 344 (2001)].

\bibitem{Buzdin_cm}
I.~Baladi\'e and A.~Buzdin, Phys. Rev. \textbf{B67}, 014523 (2003).

\bibitem{Deutscher}
G.~Deutscher and F.~Meunier, Phys. Rev. Lett. \textbf{22}, 395 (1969).

\bibitem{ANL}
J.\,Y.~Gu, C.-Y.~You, J.\,S.~Jiang et al., Phys. Rev. Lett. \textbf{89}, 267001 (2002).

\bibitem{Buzdin}
I.~Baladi\'e, A.~Buzdin, N.~Ryzhanova, and A.~Vedyayev, Phys. Rev. \textbf{B63}, 054518 (2001).

%\bibitem{gamma}
%There is a difference in the definitions of the parameter $\gamma$ between this letter and
%Refs.~\cite{Buzdin_EL,Buzdin_cm,Buzdin}. Our $\gamma$ is obtained from the one of
%Refs.~\cite{Buzdin_EL,Buzdin_cm,Buzdin} after multiplication by $\xi_s / \xi_f$.

\bibitem{JETPL}
Ya.\,V.~Fominov, N.\,M.~Chtchelkatchev, and A.\,A.~Golubov, Pis'ma Zh. Eksp. Teor. Fiz. \textbf{74}, 101
(2001) [JETP Lett. \textbf{74}, 96 (2001)].

\bibitem{PRB}
Ya.\,V.~Fominov, N.\,M.~Chtchelkatchev, and A.\,A.~Golubov, Phys. Rev. \textbf{B66}, 014507 (2002).

\bibitem{KL}
M.\,Yu.~Kupriyanov and V.\,F.~Lukichev, Zh. Eksp. Teor. Fiz. \textbf{94}, 139 (1988) [Sov. Phys. JETP
\textbf{67}, 1163 (1988)].

\bibitem{Berezinskii}
V.\,L.~Berezinskii, Pis'ma Zh. Eksp. Teor. Fiz. \textbf{20}, 628 (1974) [JETP Lett. \textbf{20}, 287
(1974)].

%\bibitem{Larkin}
%A.\,I.~Larkin, Pis'ma Zh. Eksp. Teor. Fiz. \textbf{2}, 205 (1965) [Sov. Phys. JETP Lett. \textbf{2}, 130
%(1965)].

\bibitem{reentrant}
I.\,A.~Garifullin, D.\,A.~Tikhonov, N.\,N.~Garif'yanov et al., Phys. Rev. \textbf{B66}, 020505(R) (2002).

\end{thebibliography}
\end{document}